\documentclass[%
 aip,
 amsmath,amssymb,
 reprint,%
]{revtex4-1}

\usepackage{graphicx}
\usepackage{dcolumn}
\usepackage{bm}
\usepackage{float}

\usepackage[utf8]{inputenc}
\usepackage[T1]{fontenc}
\usepackage{mathptmx}
\usepackage{etoolbox,soul}
\usepackage[normalem]{ulem}

\usepackage{color}
\newcommand{\ks}{\textcolor{black}} 

\makeatletter
\def\@email#1#2{%
 \endgroup
 \patchcmd{\titleblock@produce}
  {\frontmatter@RRAPformat}
  {\frontmatter@RRAPformat{\produce@RRAP{*#1\href{mailto:#2}{#2}}}\frontmatter@RRAPformat}
  {}{}
}%
\makeatother
\begin{document}


\title{Collision of two drops moving in the same direction}
\author{Ashwani Kumar Pal}
 \affiliation{Department of Mechanical Engineering, Indian Institute of Technology Kanpur, Kanpur - 208016, Uttar Pradesh, India}
 \author{Kirti Chandra Sahu}
 \affiliation{Department of Chemical Engineering, Indian Institute of Technology Hyderabad, Kandi - 502 284, Telangana, India}
\author{Santanu De}
\affiliation{Department of Mechanical Engineering, Indian Institute of Technology Kanpur, Kanpur - 208016, Uttar Pradesh, India}
\author{Gautam Biswas}%
\email{gtm@iitk.ac.in}
\affiliation{Department of Mechanical Engineering, Indian Institute of Technology Kanpur, Kanpur - 208016, Uttar Pradesh, India}

\date{\today}

\begin{abstract}
The collision dynamics of two drops of the same liquid moving in the same direction has been studied numerically. A wide range of radius ratios of trailing drop and leading drop ($R_r$) and the velocity ratios ($U_r$) have been deployed to understand the collision outcomes. A volume of fluid (VOF) based open-source fluid flow solver, Basilisk, has been used with its adaptive mesh refinement feature to capture the nuances of the interface morphology. The simulations are analyzed for the evolving time instances. Different collision outcomes, such as coalescence and reflexive separation with and without the formation of satellite drops, have been observed for various combinations of $U_r$ and $R_r$. The study analyzes the evolution of kinetic energy and surface energy before and after the collision for plausible outcomes. The collision outcomes are depicted on a regime map with $U_r-R_r$ space, highlighting distinct regimes formed due to variations in relevant governing parameters.
\end{abstract}

\maketitle

\section{Introduction}\label{sec:intro}

The collision of liquid drops has garnered significant attention due to its relevance in various industrial applications, including nuclear fusion, spray combustion, spray drying, inkjet printing, and design of combustion chambers \cite{biswas2021recent,kumar2020coalescence,thoroddsen2008high,stone2004engineering}. Additionally, it plays a crucial role in meteorological phenomena like raindrop formation \cite{low1982collision}. 

The Weber number influences the head-on collision of two drops, expressed as $We=\rho_l{U^2}D/\sigma$, where $\rho_l$ is the liquid density, $D$ denotes the drop diameter, and $U$ represents the relative approach velocity. As the collision unfolds, a high-pressure zone forms in the gap between the drops, causing their deformation and the extrusion of a thin surrounding fluid film. Coalescence of the drops occurs if the gap between them decreases to a scale approximately equal to the length of molecular interactions, typically of the order of $10^2$ Angstroms. Alternatively, the drops may bounce \cite{qianlaw}. \citet{qianlaw} investigated the head-on and oblique collisions of identical water and hydrocarbon drops under various surrounding pressures and identified five distinct outcomes. They are (i) coalescence after minor deformation, (ii) bouncing, (iii) coalescence after substantial deformation, (iv) coalescence followed by separation in near head-on collisions, and (v) coalescence followed by separation in off-center/ oblique collisions. The transition between the coalescence and separation regimes was also identified by \citet{qianlaw}. Subsequently, \citet{pan2008bouncing} performed numerical and experimental investigations on the head-on collision of two identical drops across a wide range of Weber numbers. Employing a time-resolved microphotographic technique similar to that of \citet{qianlaw} and a front-tracking method for numerical simulations, they found that the merging instant could be computationally assessed using an augmented van der Waals force and the associated parameter, known as the Hamaker constant. This approach was applied to collisions with both minor and significant deformations, corresponding to soft and hard collisions, respectively. \citet{nobari1996head} numerically examined the head-on collision of equal-sized droplets, utilizing a front tracking/ finite difference technique. They mapped the boundaries between coalescing and separating collisions on the Reynolds number and Weber number plane.

A few researchers have also investigated the collision dynamics of unequal-sized droplets \citep{zhang2009satellite,deka2019coalescence,chaitanya2021pof,singh2022dynamics}. \citet{zhang2009satellite} experimentally investigated the coalescence dynamics of a small drop falling onto a larger drop. \citet{kumar2020coalescence} investigated the dynamics of an ethanol drop freely descending onto a larger sessile drop of the same fluid. Their study revealed intriguing regime maps, showcasing partial coalescence and spreading behaviors, plotted against the Weber number and the volume ratio of the sessile and impacting drops. They found that the critical size ratio for satellite drops matched the conclusions of both \citet{zhang2009satellite} and \citet{nikolopoulos2012effect}, with the latter specifically examining the splashing regime triggered by drops impacting from a height. 
\citet{tang2012bouncing} theoretically and experimentally investigated the head-on collision between hydrocarbon and water drops of unequal sizes. They delineated regions in the parameter space of size ratio and collision Weber number associated with the bouncing, permanent coalescence, and separation after coalescence. On a numerical front, \citet{deka2019coalescence} explored the head-on collision of unevenly sized droplets, revealing that even with a diameter ratio of 1.2, the collision resulted in the partial coalescence of tiny daughter droplets. At high Weber numbers, \citet{cong2020numerical} scrutinized the collision dynamics of unevenly sized drops. Axisymmetric simulations conducted by \citet{goyal2020bubble} also contributed to understanding the head-on collision of two drops with disparate sizes.
Recently, attempts have also been made to study the collision of two drops having different physical properties. \citet{dirawi2020experiment} compared the collision outcomes of the drops with different viscosities with those of the same viscosities \cite{dirawi2019pof} and presented the regime maps. Numerical investigation of collision of two drops having non-equal viscosity is performed by \citet{hiranya2023}. They observed that satellite drop formation does not occur for higher viscosity ratios between the drops. \citet{paul2023investigation} investigated the dynamics of two vertically coalescing drops and a pool. They found partial coalescence when a conglomerate interacts with a pool in a manner similar to drop–pool \cite{blanchette2006partial,thoroddsen2008high,ray2010generation,kirar2020coalescence} and drop–drop \cite{zhang2009satellite,deka2019coalescence,chaitanya2021pof,singh2022dynamics} interactions. Recently, \citet{kirar2022influence} experimentally investigated the coalescence dynamics of two equal and unequal drops colliding with a deep pool in a side-by-side arrangement. They found a spectrum of coalescence outcomes, ranging from total coalescence to interacting and non-interacting partial coalescence, achieved through adjustments in drop distance and size ratios. Their study reported two distinctive coalescence phenomena, namely first, where primary drops fuse before the ensuing conglomerate integrates into the liquid pool, and another, where the drops coalesce individually within the liquid pool, giving rise to capillary wave interactions that significantly impact the overall coalescence results. \ks{The collision dynamics of a drop falling on a stationary sessile drop residing on a hydrophobic surface were experimentally and numerically investigated by \citet{sarojray2023}. They observed two distinct collision outcomes for different size ratios: reflexive separation without a satellite and reflexive separation with a satellite. A similar system was also experimentally studied by \citet{kumar2020coalescence}. They demarcated the partial coalescence and spreading behaviours in the Weber number and the volume of the sessile droplet space.} Recently, \citet{sprittles2023} presented a comprehensive review related to the bouncing of drops. In the drop-drop bouncing scenario, the bouncing dynamics can be placed in the perspective of soft transition (from merging to bouncing) and hard transition (from bouncing to merging) with increasing Weber number. \ks{An extensive review of the droplet collision dynamics can be found in \citet{amani2019numerical,liu2016numerical}.} 

As the above brief review indicates, several researchers have investigated the interaction of drops on a liquid pool, while the collision of drops receives far less attention. Furthermore, the existing studies on drop collisions have primarily focused on a head-on configuration, where the drops collide in opposite directions. However, in many practical applications and natural scenarios, drops frequently collide while moving in the same direction, an aspect examined in the current study. \ks{A few specific applications where the collision of drops moving in the same direction is observed include the collision of raindrops falling from clouds, undergoing coalescence, and the release of liquid petroleum due to an oil spill \cite{testik2011toward,lambert2016collision}.} Thus, this study also explores the effect of changing drop sizes and their direction of motion during collision events.

The rest of the manuscript is organized as follows. The details of the problem formulation, governing equations, the numerical method used and its validation are presented in \S\ref{sec:form}. The results are presented and discussed in \S\ref{sec:dis}, and the concluding remarks are given in \S\ref{sec:conc}.

\section{Formulation}\label{sec:form}
\begin{figure}[h]
\centering
\includegraphics[width=0.45\textwidth]{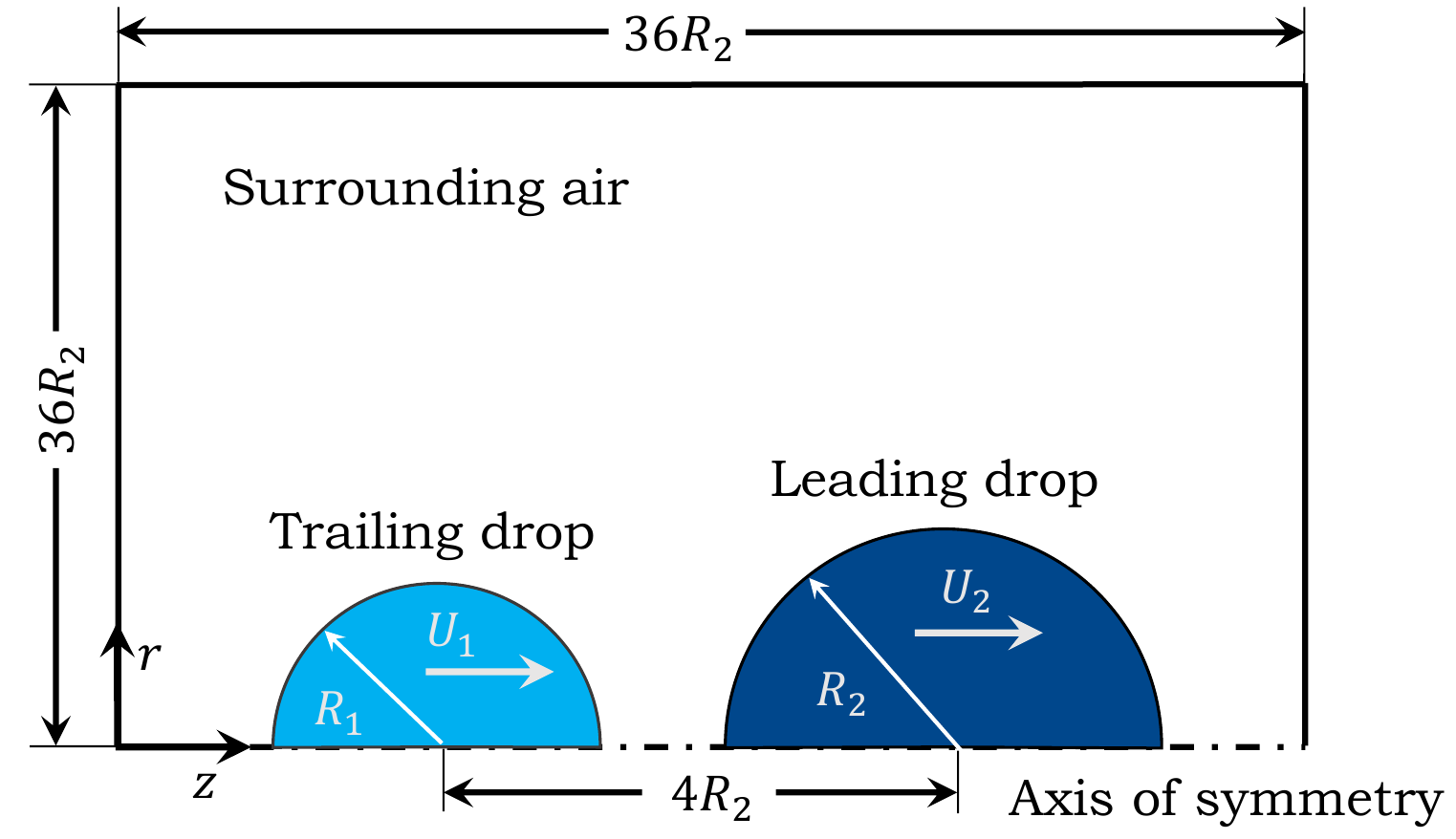}
\caption{Schematic diagram showing the collision of two drops of radii $R_1$ and $R_2$ moving with velocities $U_1$ and $U_2$ in the same direction.}
\label{fig:fig1}
\end{figure}

We investigate the collision dynamics between two drops moving in the same direction, as illustrated in Fig. \ref{fig:fig1}. The radii of the trailing and the leading drops are represented by $R_1$ and $R_2$, and they migrate with velocities $U_1$ and $U_2$, respectively, where $U_1 \ge U_2$. In our numerical simulations, the initial separation distance between the centers of the drops (at the time, $t=0$) is kept at $4R_2$. \ks{The collision dynamics of two drops moving along the coinciding axis can be effectively analyzed through axisymmetric simulations, as demonstrated in previous studies \citep{huang2019prl, hiranya2023}. However, it is essential to note that the dynamics can become three-dimensional at high velocity and size ratios when drops undergo separation, resulting in satellite drops and finger-like morphology. The understanding of the three-dimensional effect is beyond the scope of the current research.}. Thus, we employ an axisymmetric computational domain of size $36R_2 \times 36R_2$. The axisymmetric domain has a common edge coinciding with the axis of symmetry passing through the centers of the drops as enumerated in Fig. \ref{fig:fig1}. Our study considers the top part of the axis as the computational domain. The outflow boundary conditions, i.e., zero gradients for the normal and tangential components of velocity and zero Dirichlet condition for the pressure, are applied at the left and right extremes of the domain. Free slip boundary conditions are used at the top boundary, while the symmetric boundary conditions are applied at the axis of symmetry (bottom boundary).

\subsection{Governing equations} \label{subsec:eq}
The dynamics of drop collision is governed by the solution of the continuity and momentum equations, which are given by
\begin{equation}
    \nabla \cdot \textbf{u} = 0,
    \label{eqn:eq1}
\end{equation}
\begin{eqnarray}
        \rho (\phi) \left[\frac{\partial \textbf{u}}{\partial t} + \textbf{u}. \nabla \textbf{u} \right] = -\nabla p + \nabla \cdot [\mu(\phi)(\nabla \textbf{u} + \nabla \textbf{u}^T)] \nonumber\\
    + \sigma \kappa \hat{n} \delta_s + \rho(\phi)\bm{g}.
\label{eqn:eq2}
\end{eqnarray}
\ks{In Eq. (\ref{eqn:eq2}), the inclusion of the surface tension term in the Navier–Stokes equations follows the formulation proposed by \citet{brackbill1992continuum} for interfacial flows.} The velocity field is represented by $\textbf{u}$ with $u$ and $v$ being the axial and vertical components, respectively. The pressure is denoted by $p$ and $\sigma$ represents the coefficient of surface tension, $\kappa$ is the curvature of the interface, $\hat{n} $ is the unit normal vector on the interface pointing in the gas phase \ks{and $\delta_s$ represents the Dirac delta function, which is zero everywhere except at the interface.}  

An interface-capturing technique becomes essential to trace the movement of the interface within the flow accurately. The dynamics of the interface is captured by a variant of volume of fluid method available in Basilisk\citep{popinet2009}. The volume of fluid function $(\phi)$ takes a value $1$ in the pure liquid cells, $0$ in the pure gas cells, and a value between $0$ and $1$ in the cells that contain the interface. The function $\phi$ is advected using the following equation,
\begin{eqnarray}
    \frac{\partial \phi}{\partial t} + \textbf{u} \cdot \nabla \phi &=& 0.
    \label{eqn:eq3}
\end{eqnarray} 
The density $\rho(\phi)$ and viscosity $\mu(\phi)$ fields are given by 
\begin{eqnarray}
    \rho(\phi) = \rho_l \phi + \rho_g(1-\phi), \\
     \mu(\phi) ={\mu_l \phi + \mu_g(1-\phi)},
\end{eqnarray}
\ks{where $(\rho_l,\mu_l)$ and $(\rho_g,\mu_g)$ are the density and viscosity of the liquid and gas phases, respectively.}

\subsection{Numerical method} \label{subsec:numerical}
The set of governing equations is solved using an open-source code Basilisk \cite{basilisk} (\url{http://basilisk.fr}) developed by Stephane Popinet. The code uses a volume of fluid based interface capturing technique. A finite volume approach is used to discretize the governing equations on a centered grid (both pressure and velocity are defined at the center of the computational cell). \ks{Moreover, a balanced-force, continuum-surface-force formulation based on the height function has been incorporated to include the surface force term in the Navier–Stokes equations. This minimizes spurious currents at the interface.} The incompressible and variable density Navier-Stokes equations are solved using a projection method on a quadtree grid (see \citet{popinet2003, popinet2009} for a detailed procedure). A second-order accurate Bell-Colella-Glaz scheme \cite{bell-collela} is used for the advection terms in the momentum equation. The surface tension force in the momentum equation is added as the interfacial force density in the two-phase cells only. A conservative multi-dimensional scheme of \citet{weymouth2010} is used to advect the volume of fluid function. The interface is constructed geometrically as a piece-wise linear segment in the computational cell. The governing equations are marched in time using the Courant–Friedrichs–Lewy (CFL) criterion with a value of 0.5 for stable calculations. Basilisk also allows us to add as many tracers as we like. We advect a tracer to distinguish between the fluids of the two drops. \ks{The tracer follows the same advection equation as the volume of fluid function but does not influence the flow field. Its purpose is purely for visualization, facilitating the representation of distinct drops with different colors.}

\subsection{Dimensionless parameters} \label{subsec:param}

The various dimensionless numbers associated with the problem of drop collision are the Weber number $(We)$, Ohnesorge number $(Oh)$, radius ratio $(R_r)$ and velocity ratio $(U_r)$. The Weber number and the Ohnesorge number are defined based on the parameters of the leading drop as $We=2\rho_l{U_2^2}R_2/\sigma$ and $Oh=\mu_l/\sqrt{\rho_l\sigma R_2}$, respectively. \ks{The Weber number $(We)$ represents the competition between the inertia force of the liquid and the capillary force at the liquid-gas interface, while the Ohnesorge number $(Oh)$ characterizes the competition among viscous force, inertial force, and capillary forces.} The radius ratio is the ratio of the radius of the trailing drop to leading drop $(R_r=R_1/R_2)$, and the velocity ratio is the ratio of the velocity of the trailing drop to leading drop $(U_r=U_1/U_2)$. The time is non-dimensionalized as $\tau=tU_2/R_2$. \ks{The gravitational effect has a negligible effect on the collision dynamics, given the very short time scale of the coalescence process, typically on the order of milliseconds.} In the present study, we fix $R_2$, $We$, and $Oh$, and the effect of the velocity and radius ratios have been varied to obtain different collision outcomes.  

\subsection{Validation} \label{subsec:validation}

\subsubsection{Grid convergence test}
\begin{figure}[h]
\includegraphics[width=0.45\textwidth]{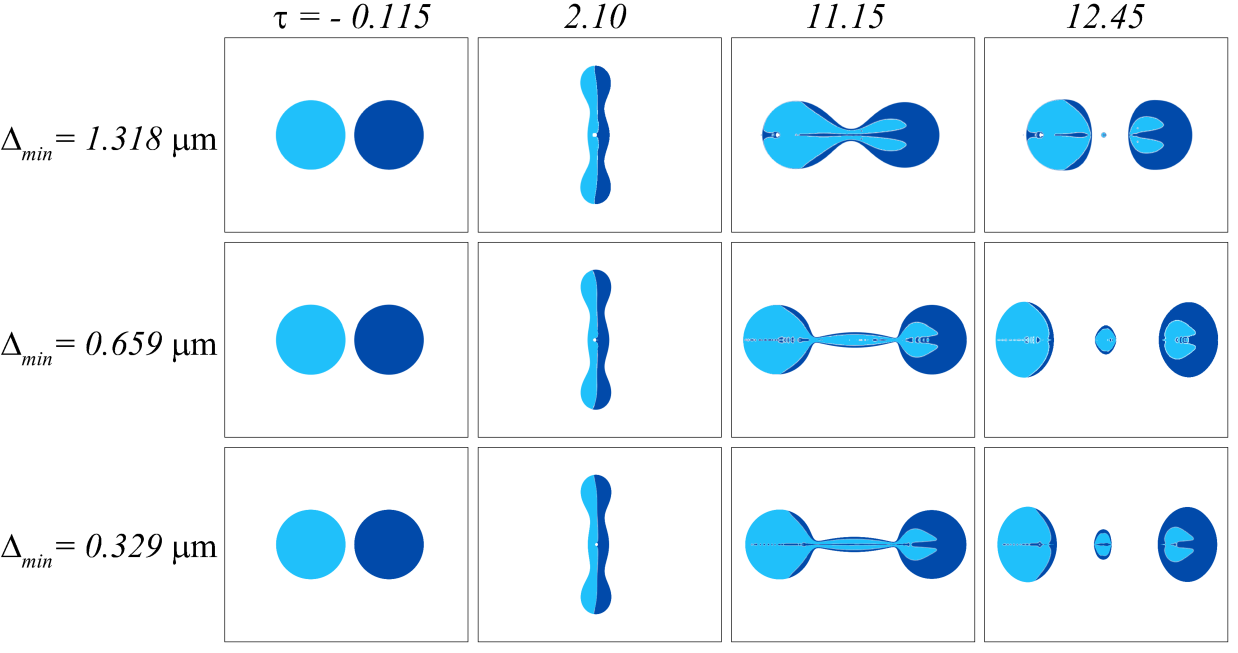}
\caption{Comparison of the drop shapes obtained using different levels of grid refinement $(L)$ at different values of the dimensionless time $\tau$, which are mentioned at the top of each panel. The parameters used for the simulation are $R_r=1.0$, $U_r=3.0$, $We=12.58$, and $Oh=0.0376$. The dimensionless time $\tau=0$ corresponds to the onset of collision. \ks{The contours of $\phi$ superimposed with the tracer function are plotted in each panel, displaying only a subset of the computational domain.}}
\label{fig:fig2}
\end{figure}

\begin{figure}[h]
\includegraphics[width=0.45\textwidth]{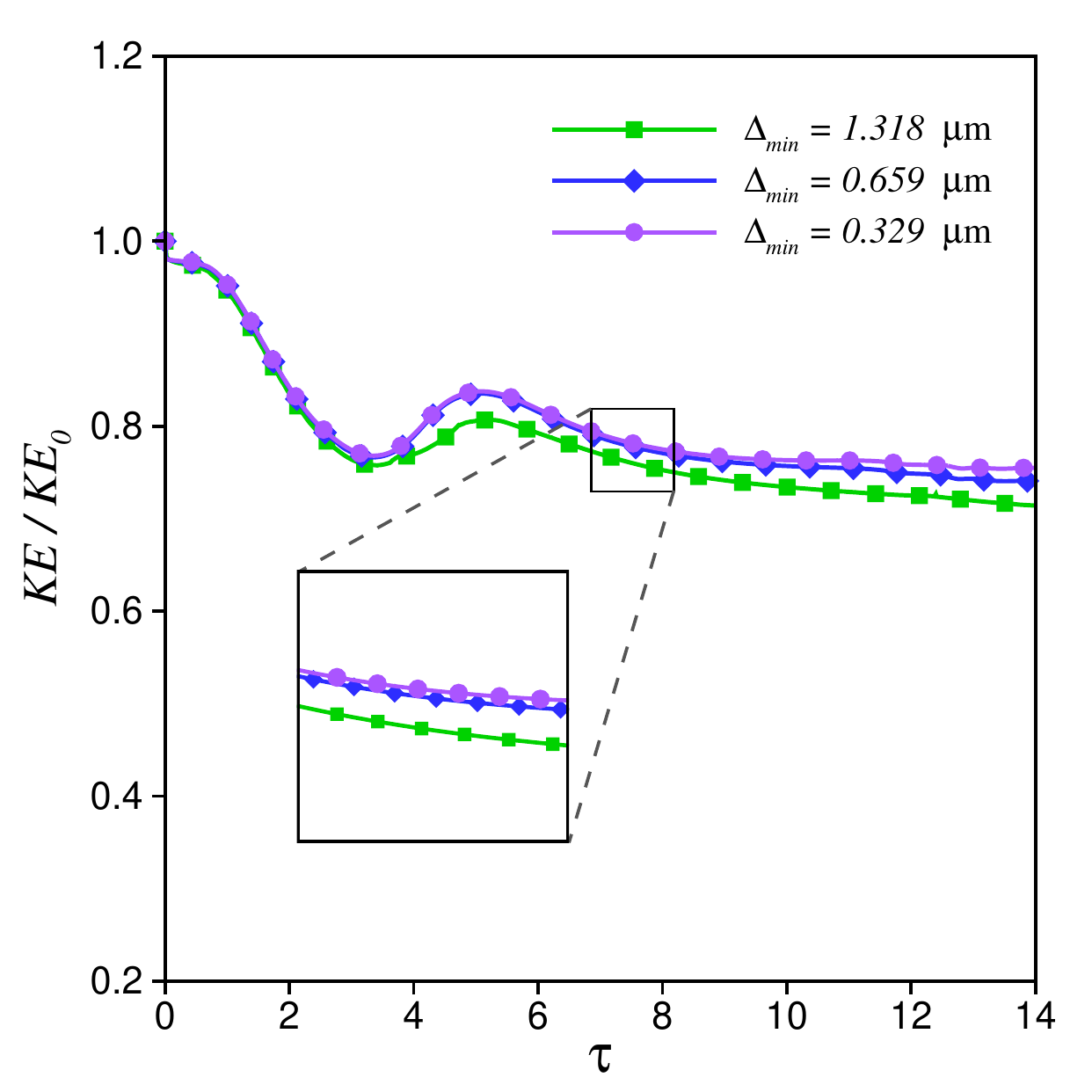}
\caption{Temporal evolution of dimensionless kinetic energy of the liquid for different levels of grid refinement. The parameters used for the simulation are $R_r=1.0$, $U_r=3.0$, $We=12.58$ and $Oh=0.0376$.}
\label{fig:fig3}
\end{figure}

The governing equations are discretized on a quadtree grid implemented in Basilisk flow solver, which makes it capable of variable grid sizes in the areas of interest. A very fine grid resolution is required to capture the physics of pinch-off in the collision process of two drops. Due to the advantage of the adaptive mesh refinement ability of Basilisk, we are able to refine the grid locally around the interface to a fine level. A grid convergence test is carried out to determine the optimum grid size that can efficiently predict the pinch-off dynamics between the drops. Two drops of identical radii $R_1=R_2=150$ $\mu$m are initialized at a distance of $4R_2$ between the centers of the drops as shown in the schematic diagram \ref{fig:fig1}. The drop collision is simulated in the reflexive separation regime in which the drops separate, giving rise to a satellite drop. The Weber and Ohnesorge numbers used for the simulations are $We=12.58$, $Oh=0.0376$. \ks{The properties the liquid and gas phases are listed in Table \ref{tab:table1}.} The grid size is represented in the form of the level of refinement, $L$. The grid size corresponding to the level of refinement $L$ can be calculated as $\Delta=(36\times R_2) / 2^L$, and thus, the grid size decreases with increasing the level of refinement. We use the grid adaptability criterion implemented in basilisk based on both the volume of fluid field $\phi$ and the velocity field $\mathbf u$ in the domain. The grid size corresponding to $L=11$ is used in the domain with an adaptive mesh refinement for the region close to the interface. Fig. \ref{fig:fig2} shows the dynamics obtained using three levels of maximum refinement, $L=12$, 13 and 14. The minimum grid sizes corresponding to these refinement levels are $\Delta_{min}=1.318$ $\mu$m, $0.659$ $\mu$m, and $0.329$ $\mu$m. It is evident from Fig. \ref{fig:fig2} that the dynamics of collision and pinch-off are almost similar for the refinement levels of $13$ and $14$, while the drops do not stretch enough for the level $12$, which makes the pinch-off asymmetric, giving rise to a relatively smaller satellite drop. \ks{The total kinetic energy of the liquid is estimated by integrating the kinetic energy of each computational cell over all the liquid cells, i.e., $KE=\Sigma { \frac{1}{2} {\rho_l F (u^2 + v^2) }dV}$, where $dV$ is the volume of the computational cell. The total surface energy is estimated as $SE=\sigma \times A_s$. In this expression, $A_s$ is the total surface area of the liquid, which is calculated by adding the interface area in all the two-phase cells.}
To make a quantitative comparison of the grid size on the simulation results, in Fig. \ref{fig:fig3}, we compare the total kinetic energy of the liquid phase obtained using different levels of refinement. The kinetic energy is non-dimensionalized with the initial kinetic energy of the liquid phase. It can be observed from Fig. \ref{fig:fig3} that a convergence of the curves is achieved as we increase the level of refinement. For optimising the computational time while maintaining decent accuracy of the computer program, the maximum refinement level of $13$ ($\Delta_{min}=0.659$  $\mu$m) is used for all the computations performed in the present study. 

\begin{table}[ht]
\caption{Properties of the fluids considered in our simulations. The surface tension $(\sigma)$ at the tetradecane-air interface $26 \times 10^{-3}$ N/m.}
\begin{ruledtabular}
\begin{tabular}{cccc}
& $\rho$ (kg/m$^3$) & $\mu$ (Pa$\cdot$s) &  \\
\hline
Tetradecane & $759.0$ & $2.05 \times 10^{-3}$  \\
Air &$1.2$ & $1.78 \times 10^{-5}$ \\
\end{tabular}
\label{tab:table1}
\end{ruledtabular}
\end{table}

\begin{figure*}
\includegraphics[width=0.8\textwidth]{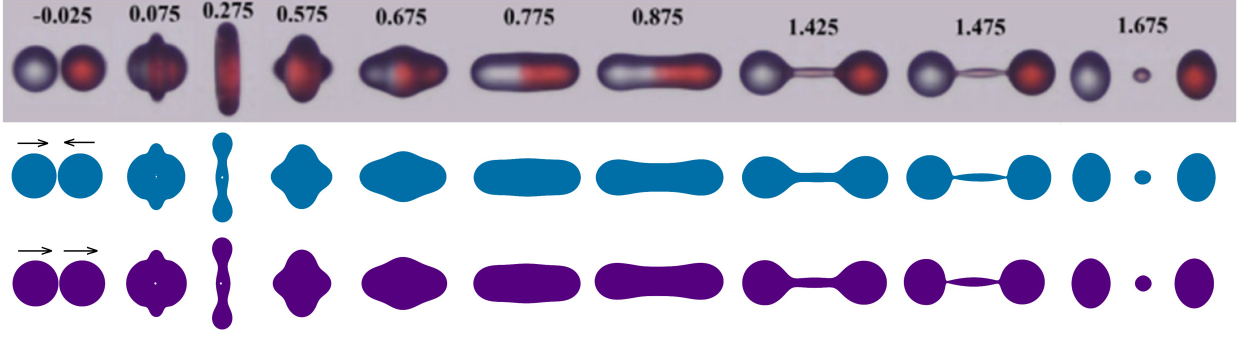}
\caption{Comparison of collision outcomes of the drops migrating in the opposite direction (head-on) and the same direction obtained from the present simulations with the results of \citet{huang2019prl}. The top panel presents the results of \citet{huang2019prl}. The middle and bottom panels show our results with the drops moving in the opposite ($U_1=U_2=1.145$ m/s) and the same ($U_1=3.046$ m/s and $U_2=0.7555$ m/s) directions, respectively. Thus, the relative velocity for both cases remains $U=2.29$ m/s. The rest of the parameters used for the simulations are $R_r=1.0$, $We=11.48$, and $Oh=0.0376$.}
\label{fig:fig4}
\end{figure*}

\begin{figure*}
    \centering
    \includegraphics[width=0.8\textwidth]{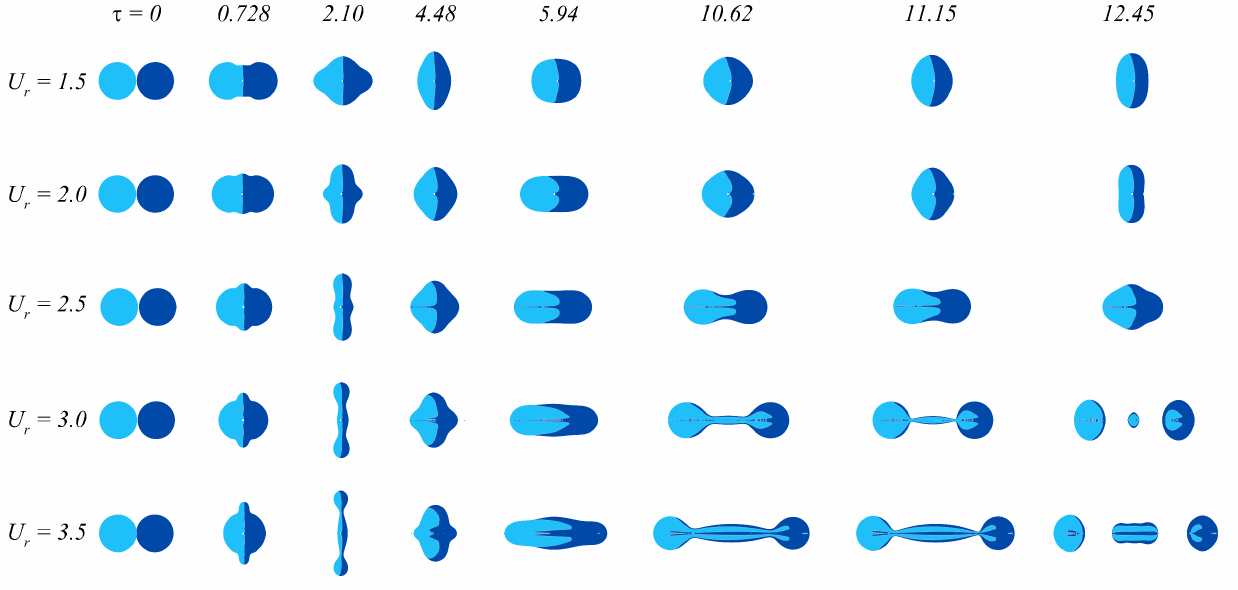}
    \caption{Evolution of the shapes of the drops with $R_r=1.0$ undergoing collision for different values of velocity ratio ($U_r$). The parameters used for the simulation are $We=12.58$ and $Oh=0.0376$. The dimensionless time $\tau=0$ corresponds to the onset of collision.}
    \label{fig:fig5}
\end{figure*}
\subsubsection{Comparison with the results of \citet{huang2019prl}.}
To check the numerical accuracy of the results obtained in the present simulations, a case of two identical tetradecane drops colliding in the opposite direction is considered. The results are compared with those obtained by \citet{huang2019prl}. Two tetradecane drops with identical radius $R_1=R_2=150$ $\mu$m, and initial velocities $U_1=U_2=1.145$ m/s in the opposite directions are simulated in an axisymmetric configuration. The Weber number and Ohnesorge number for the simulation are kept same as by \citet{huang2019prl} ($We=11.48$ and $Oh=0.0376$). The collision outcomes are plotted with the results of \citet{huang2019prl} at different time instances in Fig. \ref{fig:fig4}. The results obtained from the present simulations show satisfactory agreement with the results of the \citet{huang2019prl}. As observed by them, the drops undergo symmetrical reflexive separation after the collision and produce a small satellite drop in between two larger drops. This verifies the accuracy of the results obtained from the present simulations.

\section{Results and discussion}\label{sec:dis}

We begin the presentation of our results by analysing the dynamics of two drops moving in the same direction. Simulations are performed to see the effect of velocity and radius ratios between the drops on the collision outcome. The velocity, $U_2=1.15$ m/s, and the radius, $R_2=150$ $\mu$m of the leading drops are kept fixed for all the cases presented in this paper. Thus, the dimensionless parameters, such as $We=12.58$ and $Oh=0.0376$ are constant. The radius $R_1$ and velocity $U_1$ of the trailing drop are varied to obtain various radius and velocity ratios considered. The collision process exhibits three distinct outcomes: coalescence and reflexive separation with and without satellite drop formation. The transition from coalescence to reflexive separation takes place as the total inertia of the liquid increases for the higher values of $U_r$ (\citet{qianlaw}). However, the formation of the satellite drop is inhibited in the case of higher radius ratio $R_r$ as also observed by \citet{huang2019prl}.

\begin{figure*}
    \centering
    \includegraphics[width=0.8\textwidth]{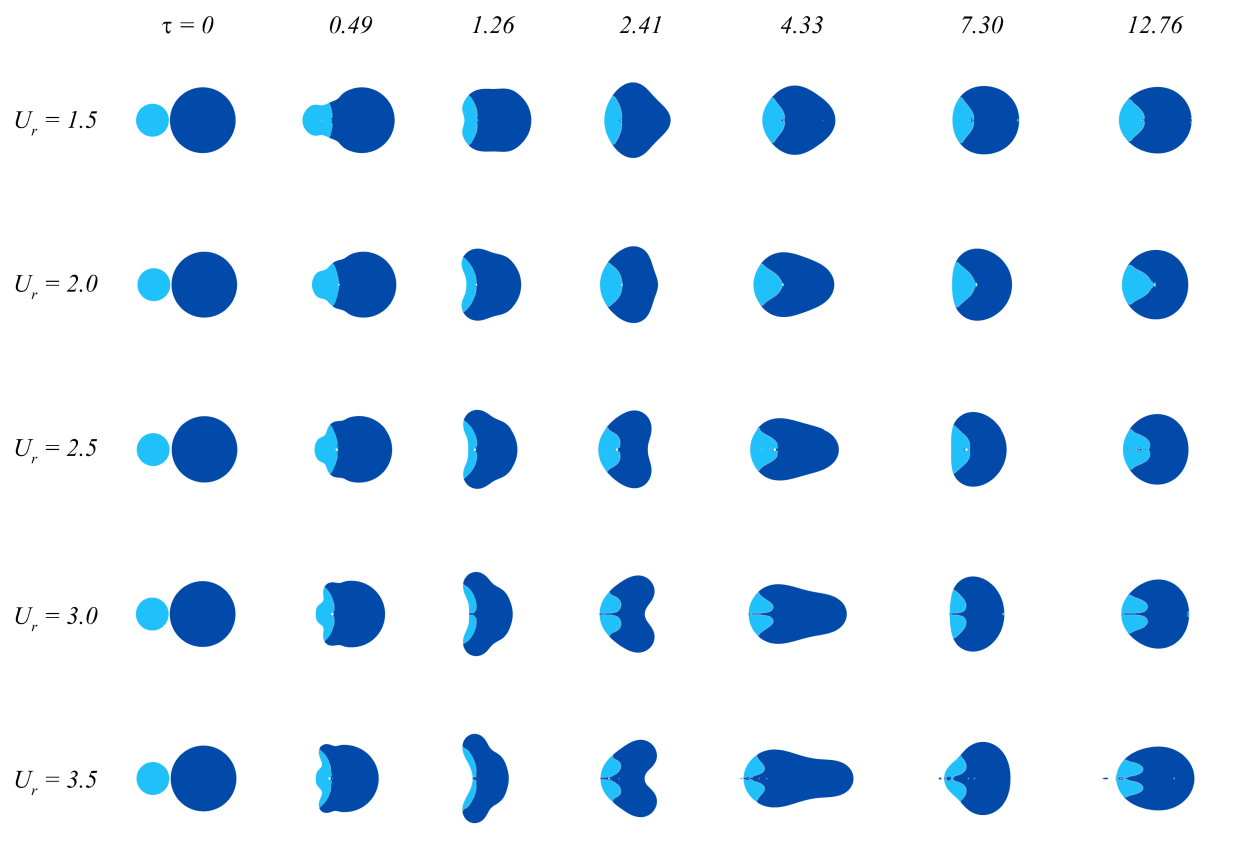}
    \caption{Evolution of the shapes of the drops with $R_r=0.5$ undergoing collision for different values of velocity ratio. The parameters used for the simulation are $We=12.58$ and $Oh=0.0376$. The dimensionless time $\tau=0$ corresponds to the onset of collision.}
    \label{fig:fig6}
\end{figure*}

\subsection{Effect of velocity ratio for identical drops}

In this section, we investigate the effect of velocity ratio on the collision dynamics of two identical drops. Two tetradecane drops of the same size ($R_1=R_2=150$ $\mu$m) are given the initial velocities $U_1$  and $U_2$ in the same direction ($U_1 > U_2$). The numerical configuration and the boundary conditions remain the same, as explained in the formulation section \ref{sec:form}. Simulations are performed for various values of the velocity ratio. The trailing drop approaches the leading drop with a relatively higher velocity as the velocity ratio increases with the velocity of the leading drop fixed for all the cases. The shapes of the drops are plotted in Fig. \ref{fig:fig5} at different time instances for different values of the velocity ratios considered. The time is non-dimensionalized by the time constant of the leading drop ($\tau = tU_2/R_2$), and $\tau = 0$ represents the moment when the two drops are about to merge. Top-to-bottom panels show the evolution of collision profiles with increasing the velocity ratio. Visual inspection of Fig. \ref{fig:fig5} shows the change of outcome regime of the collision from coalescence to reflexive separation. It can be observed that the drops come close to each other, forming a single liquid bulk that keeps moving in the direction of the velocities of the drops. 

In the case of lower velocity ratios (up to $U_r=2.5$), the combined liquid mass stretches out in the vertical direction and then in the horizontal direction. These oscillations keep happening while the liquid mass moves in the horizontal direction, as seen in the coalescence of two drops moving in the opposite direction \citet{qianlaw}. During this process, the conversion of surface energy into kinetic energy and vice versa occurs until the shape fluctuations are damped due to viscous damping and the liquid mass takes a spherical shape, which moves in the direction of initial velocities of the drops. However, in the case of higher velocity ratios ($U_r \ge 3.0)$, the drops undergo separation after the collision. This happens due to increased relative impact velocity between the drops \citet{huang2019prl}. The trailing drop impacts the leading drop with high velocity, imparting momentum to the trailing drop. After the collision, the liquid packet squeezes, forming a thin disk shape, which starts to elongate along the axis due to the action of the capillary forces. The combined liquid mass starts retracting and elongating in the axial direction. Since the liquid mass has higher momentum than in the case of lower velocity ratios, it keeps on elongating and forms a dumbbell shape with a ligament connecting two liquid masses at its end. The liquid mass keeps moving in the axial direction during this process. The pressure starts increasing at the two ends of the ligament, which carries the fluid away from the ends \citet{hiranya2023}. This gives rise to a symmetrical pinch-off of the ligament, and the liquid inside the ligament gives rise to a satellite drop. The leading part of the ligament detaches with a higher velocity than the one on the trailing while the liquid continues to move in the same direction. A significant increase in the size of the satellite drop is also observed in the case of $U_r=3.5$ as compared to the case with $U_r=3.0$.

\begin{figure*}
    \centering
    \includegraphics[width=0.8\textwidth]{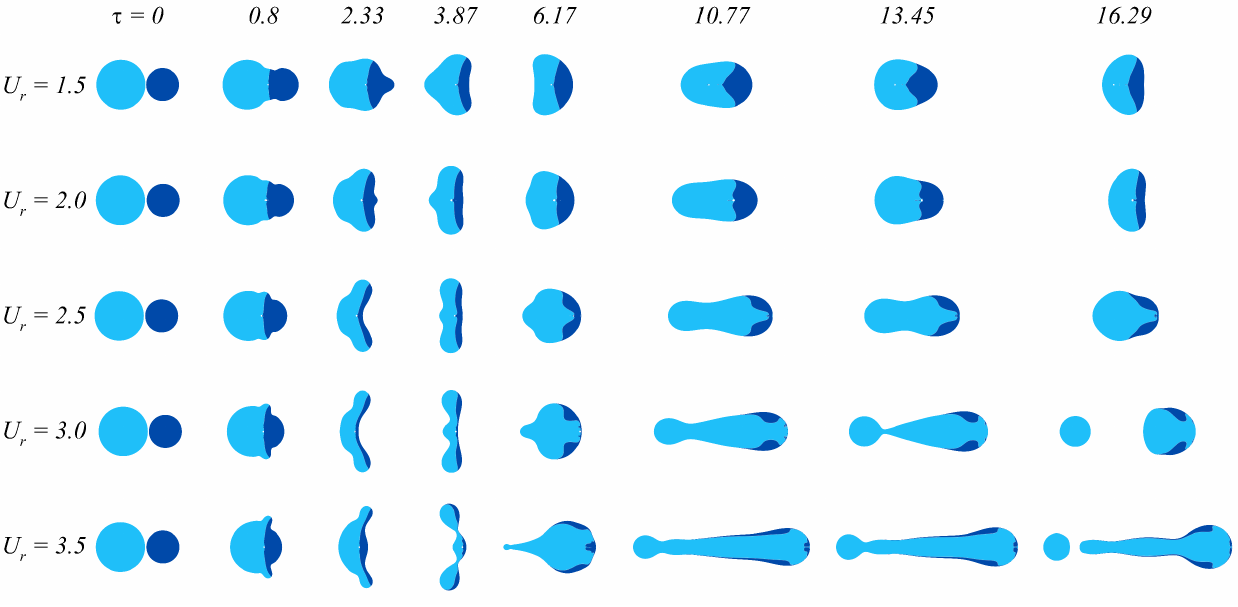}
    \caption{Evolution of the shapes of the drops with $R_r=1.5$ undergoing collision for different values of velocity ratio. The parameters used for the simulation are $We=12.58$ and $Oh=0.0376$. The dimensionless time $\tau=0$ corresponds to the onset of collision.}
    \label{fig:fig7}
\end{figure*}

\subsection{Effect of velocity ratio for the different sized drops}

The effect of the size of the drops, along with the velocity ratio, on the collision outcome of the drops is analyzed in this section. The radius $(R_1)$ and velocity $(U_1)$ of the trailing drop are varied by keeping the radius $(R_2)$ and velocity $(U_2)$ of the leading drop fixed to obtain various radius and velocity ratios. The simulations are performed for various combinations of radius and velocity ratios.

Firstly, we extend the analysis for the cases with $R_r<1$. For the case of $R_r<1$, the \ks{trailing} drop is initialized with a radius $R_1 = 0.5  R_2$ and with different initial velocities $(1.5  U_2 \leq U_1 \leq 3.5 U_2)$. The collision outcomes for different values of the velocity ratios are plotted in Fig. \ref{fig:fig6}. The trailing drop, which is smaller, hits the larger leading drop with higher velocity. The collision in these cases is asymmetric, unlike the previous case with identical-sized drops. The smaller drop penetrates the larger drop while the larger absorbs the higher velocity impact after collision. The impact creates a crater at the trailing side of the combined liquid mass while the leading side remains convex. The curvature of the crater increases with increasing the velocity ratio of impact, as expected. As a consequence of high-velocity impact, strong clockwise vortexes are generated towards the ends of the crater, which rotates the liquid mass in such a way that the leading side of the liquid takes a concave crater shape and the trailing side becomes convex shaped for higher velocity ratios ($U_r>2.0$). In contrast, such rotation of the ends of the crater does not take place for lower values of the velocity ratio. At later times, as evident from Fig. \ref{fig:fig6}, the liquid packet starts to stretch from its leading end for all the cases of the velocity ratios, which is also the case in the case of collision of identical-sized drops. The stretching is more significant in the case of collision with higher velocity ratios, while it is not that prevalent in the case of lower velocity ratios where the liquid packet takes an almost spherical shape due to capillary action and lack of momentum. Even in the case of higher velocity ratios, the liquid eventually takes a practically spherical shape due to containing enough momentum to separate the liquid packet into two parts. Formation of the satellite drops is not observed for any values of the velocity ratio considered in the present study. The liquid continues to migrate in the direction of initial velocity with an almost spherical shape.

\begin{figure*}
    \centering
    \includegraphics[width=0.8\textwidth]{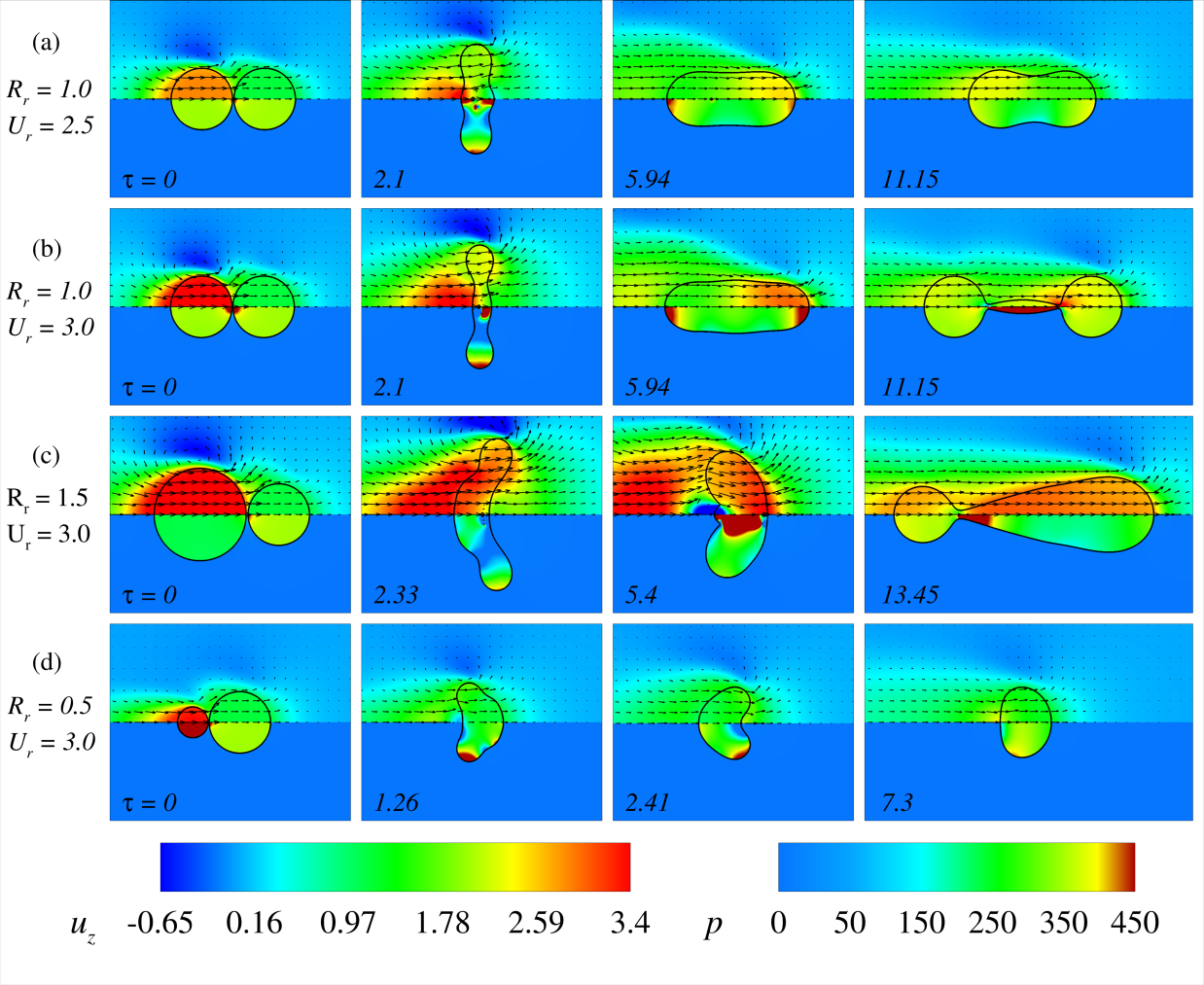}
    \caption{Contours of the axial velocity (upper half) and pressure field (bottom half) along with the interface of the drops for different combinations of $U_r$ and $R_r$. The top to bottom rows represent the coalescence, reflexive separation with a satellite drop, and reflexive separation without a satellite, respectively. The rest of the dimensionless parameters used for the simulation are $We=12.58$ and $Oh=0.0376$. }
    \label{fig:fig8}
\end{figure*}

Next, we study the effect of radius ratio $R_r>1.0$. The trailing drop is initialized with a radius higher than the leading drop and with different initial velocities, giving rise to various velocity ratios. The outcomes of the collision process for different velocity ratios for $R_r=1.5$ are shown in Fig. \ref{fig:fig7}. The larger trailing drop hits the smaller leading drop with high impact velocities. As a result, for these cases, the crater is formed on the leading side of the impacting drop due to less momentum of the leading drop. The leading drop fluid is spread on the outer side of the crater. For higher velocity ratios, the drop tries to attain a disk shape before the ends are rotated to reverse the direction of the crater. The combined liquid packet starts to elongate as time progresses. The stretching of the liquid is asymmetric in these cases, and the trailing side of the liquid is stretched more than the leading one. The stretching is continued until the relative kinetic energy is converted into the surface energy or dissipated due to viscous effects. For lower values of the velocity ratios ($U_r<2.5$), the liquid does not break into the part due to lack of momentum, and the capillary forces pull back the liquid. The liquid packet again tries to extend itself in the axial depending upon the momentum and finally attains a stable shape. In the case of higher values of velocity ratios ($U_r>2.5$), the liquid breaks into two parts since it has more momentum to overcome the capillary and viscous forces. Unlike the case with the collision of identical drops, the stretching of the liquid volume is asymmetrical in this case, which leads to an asymmetrical break-up of the liquid volume. The necking is only at one location towards the trailing end where the pressure is maximum, and the pinch-off takes place at this location. The liquid breaks into two unequal volumes of liquid, the trailing being the smaller one and the leading one being the larger one. The separated liquid volumes try to attain a stable spherical shape while moving in the direction of their initial velocity.

\begin{figure*}
    \centering
    \includegraphics[width=0.8\textwidth]{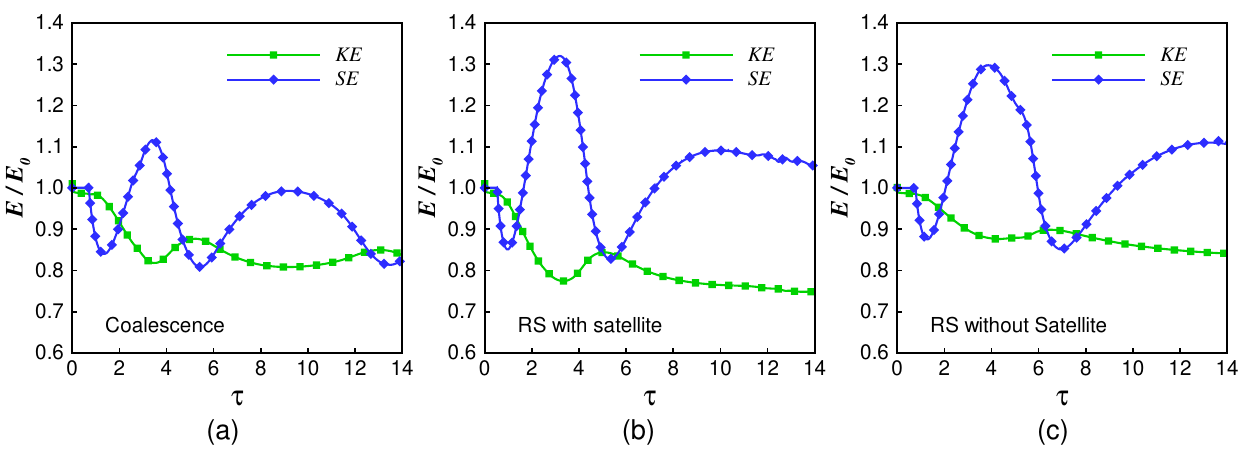}
    \caption{Evolution of the normalised surface energy and kinetic energy for different collision outcomes: (a) coalescence for $R_r=1.0$ and $U_r=2.5$, (b) reflexive separation with satellite for $R_r=1.0$ and $U_r=3.0$, (c) reflexive separation without satellite for $R_r=1.5$ and $U_r=3.0$. The other parameters used for the simulation are $We=12.58$ and $Oh=0.0376$. }
    \label{fig:fig9}
\end{figure*}

\subsection{Different collision outcomes and phase diagram in $R_r-U_r$ plane}

Various collision outcomes observed in the present computations are due to the different values of $U_r$ and $R_r$. We analyze the evolution of velocity and pressure contours for these cases to understand the mechanism behind different collision outcomes. In Fig. \ref{fig:fig8}, the contours of the axial velocity (top half) superimposed with the velocity vectors and the pressure contours (bottom half) are shown for four combinations of $U_r$ and $R_r$. These combinations cover the different cases of collision outcomes observed in the present study. The contours are plotted at various time instances to observe the evolution of the velocity and pressure fields as the collisions occur. 

Figures \ref{fig:fig8}a and \ref{fig:fig8}b show the contours for the collision of two identical sized drops ($R_r=1.0)$ with $U_r=2.5$ and $U_r=3.0$ respectively. The drops undergo coalescence with no separation and reflexive separation with satellite formation for $U_r=2.5$ and $U_r=3.0$, respectively. For both cases, the trailing drops approach the leading drops with high velocities while the pressure remains the same in both drops (due to the same size) before the collision. The velocity circulating zones are observed close to the trailing drops due to their high velocities, while no such circulation occurs near the leading drops. High-pressure zones are created near the point of impact inside the drops as they come close to each other, giving rise to the pinch-off of the interface at these points.

For both these cases, the liquid takes a thin disk type of shape after collision (see second panels of Figs. \ref{fig:fig8}a and \ref{fig:fig8}b). Due to high pressure at the ends of disk-shaped liquid packet, the ends are stretched in the radial direction. The pressure at the ends is higher for the case of $U_r=3.0$; thus, the liquid stretches more, making the disk thinner. The liquid is pulled back towards the axis of the disk due to capillary pull, and it starts to elongate in the axial direction. The pressure becomes high at the two ends in the axial direction, stretching the liquid packet in the axial direction. For the case of $U_r=2.5$, the pressure is not high enough to pull the two ends apart, and the liquid is pulled back due to the action of capillary forces. Thus, the liquid packet remains moving in the axial direction as a whole. In the case of $U_r=3.0$, the pressure at the ends is high enough to pull the liquid mass, making it a dumbbell shape where a thin ligament connects two blobs of liquid. The pressure at the ends of the ligament is higher than the liquid blobs, and thus, the pinch-off takes place at these two ends of the ligament, giving rise to a small size satellite drop. 

The contours of velocity and pressure for the collisions of unequal-sized drops are shown in Figs. \ref{fig:fig8}c and \ref{fig:fig8}d. For both these cases, since $U_r>1$, the trailing drops approach the leading drops with high velocities. The pressure at the point of impact increases as the drop interfaces come close to each other. Unlike the case of equal-size collisions, the collisions occur with different pressures in the leading and trailing drops. The \ks{pressure} of the leading drop is relatively higher than the trailing drop for the case of $R_r=1.5$ (see panel-1 of Fig. \ref{fig:fig8}c) due to its smaller size. While it is higher in the trailing drop than the leading drop for the case of a collision with $R_r=0.5$ (see panel-1 of Fig. \ref{fig:fig8}d) for the same apparent reason. The collision of the high-pressure drop with the low-pressure drop forms a crater-type structure on either of the two sides of the liquid bulk after the collision. The side of the high-pressure drop (smaller sized) decides on which side the crater is to be formed (on the leading side for $R_r=1.5$ and on the trailing side for $R_r=0.5$). Intense velocity circulations with high-pressure zones are present at the ends of the bow-shaped liquid structure, which try to bend the bow shape and reverse the side on which the crater is present. Higher \ks{ pressure} at the ends pushes the liquid toward the center, and the shapes start to elongate in the axial direction. For the case of $R_r=0.5$, the high velocity trailing drop does not transfer enough momentum to the liquid bulk, and thus, the axial elongation is not high enough to overcome the capillary force to separate the liquid into two parts. So, the liquid shape oscillates in the radial and axial directions while moving in the axial direction. However, in the case of $R_r=1.5$, the high velocity and bigger size drop transfers large momentum to the liquid bulk on collision. An asymmetric high-pressure zone is present near the axis of the bow-shaped due to the difference in the curvature of the interface at the leading and the trailing side of the liquid bulk. This high-pressure zone pushes the liquid out from the trailing side of the bow while elongating in the axial direction. The neck formation starts towards the trailing side of the center due to the action of the capillary force (panel-4 of Fig. \ref{fig:fig8}c). The pressure in the necked region increases, and the fluid flows away from the neck \citep{hiranya2023} and the pinch-off takes place at the necking point, dividing the liquid mass into two parts, the trailing one being the smaller and the leading one being the larger one.

\begin{figure}
    \centering
    \includegraphics[width=0.45\textwidth]{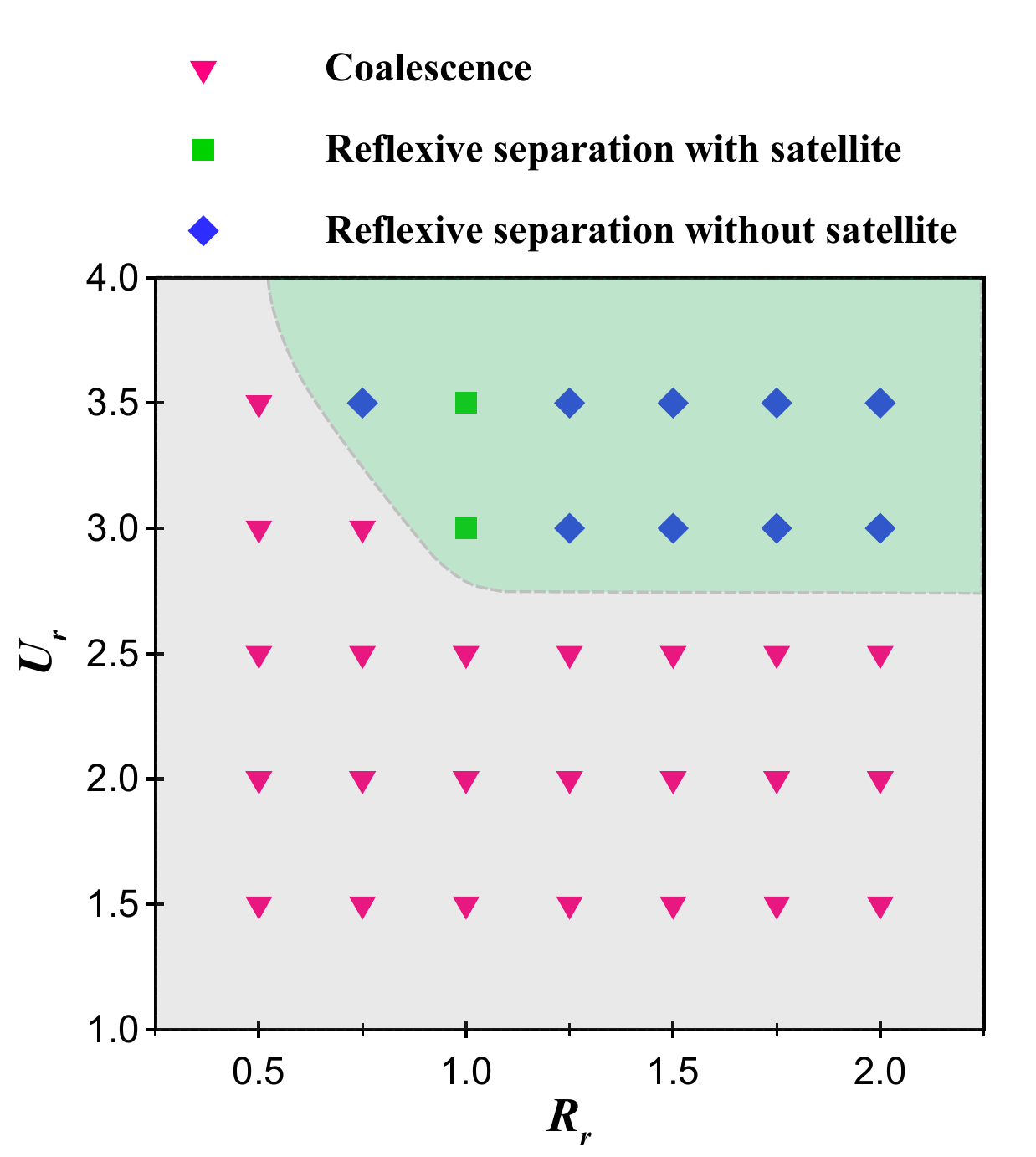}
    \caption{Phase diagram demarcating different collision outcomes, such as coalescence and reflective separation with and without satellite drop in $U_r-R_r$ space.}
    \label{fig:fig10}
\end{figure}

In Fig. \ref{fig:fig9}, the kinetic and surface energy time variations are plotted during the collision process for various collision outcomes. The energy curves are presented for three cases with distinct collision outcomes, i.e., coalescence (Fig. \ref{fig:fig9}a), reflexive separation with satellite (Fig. \ref{fig:fig9}b), and reflexive separation without satellite formation (Fig. \ref{fig:fig9}c). The kinetic and surface energy presented in Fig. \ref{fig:fig9} are normalized with the total initial kinetic and surface energy of the drops, respectively. In the binary collision of drops, there is a competition between inertial and capillary forces \citep{pan2008bouncing}. \ks{During the collision process, the capillary action tries to minimize the surface energy of the liquid while the inertial forces try to extend the liquid interface by working against them.
The combined effect of both these forces is responsible for deciding the liquid shapes at different stages of collision. As the two drops approach each other, the surrounding fluid between the drops drains out, and coalescence of the interfaces occurs. For all three distinct cases of drop collision, the surface energy decreases and attains a minimum value due to the combined effect of inertial and capillary forces after collision. The surface energy starts increasing as the liquid packet deforms in the axial direction and extends in the radial direction due to the effect of inertia. The surface energy exhibits maxima corresponding to a time when the liquid has its maximum axial deformation and radial expansion, taking a thin disk shape. The capillary forces try to minimize the surface energy, and the disk is retracted back, extending the liquid in the axial direction and causing a local minimum to occur again. Due to inertial effects, the surface energy increases as the liquid grows further in the axial direction. For coalescence, this process of increase and decrease in the surface energy is repeated for several cycles unless all the velocity fluctuations are damped due to viscosity and the combined liquid takes the shape of a spherical drop with minimum surface energy. However, in the case of reflexive separation (both with and without satellite formation), no such oscillations in the surface energy are observed. The total surface energy becomes almost constant after the reflexive separation of the drops. For all three distinct collision outcomes, the total kinetic energy of the liquid decreases as the two drops coalesce with each other, and the axial deformation of the fluid starts. It takes the minimum value when the two drops are in maximum axial deformation and maximum radial expansion (the kinetic is stored as the surface energy of the liquid). The kinetic energy recovery from surface energy occurs as the liquid stretches in the axial direction and attains a local maximum value corresponding to the time when the liquid has local minimum surface energy. In the case of coalescing drops, the total kinetic energy of liquid oscillates in opposite phases, with the surface energy showing the competition between the two kinds of energy. However, in the cases of reflexive separations, the kinetic energy steadily decreases after attaining a local maximum value. The decrease in the kinetic energy after impact and its recovery during the axial extension of the liquid is higher in the case of reflexive separation with satellite formation than in the case of separation without satellite (see Figs. \ref{fig:fig9}b and \ref{fig:fig9}c). This happens because in the case of $R_r=1.5$, the trailing drop experiences less inertial resistance from the small size of the leading drop than in the case of equally sized leading drop for $R_r=1.0$.} 

As discussed earlier, the collision outcome of two drops moving in the same direction depends on the collision parameters such as radius ratio $R_r$ and velocity ratio ($U_r$) between them. Simulations with several combinations of $R_r$ and $U_r$ are performed to observe the different collision outcomes. In Fig. \ref{fig:fig10}, a $R_r - U_r$ regime map is plotted to distinguish between various collision outcome regimes. It is noted that the formation of satellite drops takes place only for the collision of identical-sized drops ($R_r=1.0$) with higher values of velocity ratios ($U_r>2.5$) (where the liquid has enough inertia to overcome the capillary forces and extend the interface into multiple parts).

\section{Concluding remarks}\label{sec:conc}

We have numerically examined the collision dynamics of two drops moving in the same direction. The relevant collision parameters, namely the velocity and radius ratios, are varied to observe various collision outcomes. Three distinct collision outcomes observed in the present study are coalescence, reflexive separation with satellite, and reflexive separation without satellite formation. The collision dynamics is explained with the help of the pressure and velocity contours. The evolution of kinetic and surface energy for different cases of collisions is also presented. For the collision of identical-sized drops ($R_r=1.0$), the drops undergo coalescence for $U_r \leq 2.5$, while reflexive separation with satellite takes place for $U_r>2.5$. Similarly, in the case of $R_r=1.5$, coalescence occurs for $U_r \leq 2.5$, and reflexive separation without formation of satellite drop (asymmetric reflexive separation) takes place for $U_r>2.5$. However, for the cases of $R_r=0.5$, the collisions exhibit coalescence for all the values of velocity ratios considered in the present study. \ks{The temporal evolution of normalized kinetic and surface energy of the liquid is depicted for the three distinct collision outcomes. It is evident that the liquid's kinetic energy is converted into surface energy during axial deformation and subsequently released during axial expansion. In the case of coalescence, oscillations of opposite phases are observed in both kinetic and surface energy until they are dissipated by viscous effects. Such oscillations are not evident in drop collisions leading to separation.} 
Finally, through a large number of simulations, the collision outcomes have been summarized, and the boundary separating coalescence and reflexive separations is identified.

\section*{Acknowledgements:} A.K.P. acknowledges the support from PMRF fellowship, Government of India and G.B. acknowledges his gratitude to J. C. Bose National Fellowship of SERB, Government of India (grant: JBR/2020/000042). K.C.S. thanks the Science and Engineering Research Board, India for the financial support through grant CRG/2020/000507. The authors express their sincere appreciation to the high-performance computing facilities (HPC and Paramsanganak) of IIT Kanpur for providing computational resources.

\vspace{-0.5cm}

\section*{Data Availability Statement}
The data that support the findings of this study are available from the corresponding author upon reasonable request.

\section*{REFERENCES}

%

\end{document}